\documentclass[twocolumn,aps,prd,amsmath,amssymb,nofootinbib,superscriptaddress]{revtex4-1}\newcommand{\cw}{\columnwidth}\newcommand{\preprintnumber}{\hfill MIT-CTP 4625\,\,\,\,\maketitle}

\usepackage{graphicx,bm}
\usepackage{url}

\newcommand{\beq}{\begin{equation}}
\newcommand{\bea}{\begin{eqnarray}}
\newcommand{\eeq}{\end{equation}}
\newcommand{\eea}{\end{eqnarray}}

\newcommand{\mpl}{M_{Pl}}
\newcommand{\dPsi}{\delta\Psi}
\newcommand{\gam}{\gamma}
\newcommand{\kuv}{k_{\mbox{\tiny{UV}}}}

\begin{document}

\title{Do Dark Matter Axions Form a Condensate with Long-Range Correlation?}

\author{Alan H.~Guth}
\email{guth@ctp.mit.edu}
\affiliation{Center for Theoretical Physics \& Department~of Physics,\\ Massachusetts Institute of Technology, Cambridge, MA 02139, USA}

\author{Mark P.~Hertzberg}
\email{mphertz@mit.edu}
\affiliation{Center for Theoretical Physics \& Department~of Physics,\\ Massachusetts Institute of Technology, Cambridge, MA 02139, USA}
\affiliation{Institute of Cosmology \& Department of Physics and Astronomy, \\Tufts University,
	Medford, Massachusetts 02155, USA}

\author{C.~Prescod-Weinstein}
\email{chanda@mit.edu}
\affiliation{MIT Kavli Institute for Astrophysics and Space Research \& Dept.~of Physics,\\ Massachusetts Institute of Technology, Cambridge, MA 02139, USA}

\date{\today}

\begin{abstract}
Recently there has been significant interest in the claim that dark matter axions gravitationally thermalize and form a Bose-Einstein condensate with cosmologically long-range correlation. This has potential consequences for galactic scale observations.
Here we critically examine this claim. We point out that there is an essential difference between the thermalization and formation of a condensate due to {\em repulsive} interactions, which can indeed drive long-range order, and that due to {\em attractive} interactions, which can lead to localized Bose clumps (stars or solitons) that only exhibit short range correlation. 
While the difference between repulsion and attraction is not present in the standard collisional Boltzmann equation, we argue
that it is essential to the field theory dynamics, and we explain why the latter analysis is appropriate for a condensate. Since the axion is primarily governed by {\em attractive} interactions -- gravitation and scalar-scalar contact interactions -- we conclude that while a Bose-Einstein condensate is formed, the claim of long-range correlation is unjustified.
\end{abstract}

\preprintnumber

%\maketitle
%\newpage
\tableofcontents

\section{Introduction}

Cosmological observations, such as galaxy rotation curves and anisotropies in the cosmic microwave background radiation, indicate that the majority of matter in the universe is a non-radiating type known as dark matter \cite{Peebles:2013hla}. Dark matter appears to make up around five times more mass than ordinary matter, yet we know very little about its properties. Observational constraints indicate that dark matter is non-baryonic, cold and collisionless in nature, a picture known as cold dark matter \cite{2014A&A...571A..16P}. It is important to develop dark matter models with clear signatures.

Several candidates for the dark matter particle have been proposed, including weakly interacting massive particles, sterile neutrinos, and axions, among others. The latter is a hypothesized particle introduced to solve the CP problem in QCD \cite{Peccei:1977hh,Weinberg:1977ma,Wilczek:1977pj}. This particle physics motivation for axions make them a theoretically attractive candidate. In addition, the proposed mass range and non-relativistic behavior are fitting for the dark matter problem. 

Axion dark matter has a rich history, including computations that show the axion can plausibly carry the right dark matter abundance, e.g., see Refs.~\cite{Preskill:1982cy,Dine:1982ah,Abbott:1982af,Kim:2008hd}.
Such axion dark matter is currently being explored in interesting table top experiments, such as ADMX \cite{Asztalos:2009yp,Hoskins:2011iv}, utilizing the axion to photon coupling, which is a unique signature (other proposed search strategies include Refs.~\cite{Roberts:2014dda,Sikivie:2014lha}). Furthermore, axions in an inflationary cosmology can generate interesting isocurvature signatures \cite{Fox:2004kb,Beltran:2006sq,Hertzberg:2008wr}, and various other interesting ideas include Refs.~\cite{Turner:1990uz,Kawasaki:2013ae,Hertzberg:2012zc,Cheung:2011mg,Khlopov:1999tm}. Here we examine a fascinating new proposal for a cosmological or galactic scale signature of the axions, which is deeply intertwined with their bosonic character. 

Axions are essentially non-relativistic with an approximately conserved particle number, and are produced at high occupancy.
Thus they have the capacity to form a Bose-Einstein condensate (BEC). Recently, it has been proposed that axionic dark matter will gravitationally thermalize and form a BEC during the radiation dominated era \cite{Sikivie:2009qn,Erken:2011dz}. It is then argued that this causes the axion field's correlation length to grow dramatically, becoming an appreciable fraction of the size of the horizon. Furthermore, it is claimed that this produces a unique signature of $\sim 10$\,kpc caustics with a ring geometry in galaxies \cite{2013PhRvD..88l3517B}. There have been many followup studies of this fascinating idea including Refs.~\cite{Davidson:2013aba,deVega:2014wya,Noumi:2013zga,Davidson:2014hfa} and similar but distinct ideas such as Refs.~\cite{RindlerDaller:2011kx,Chavanis:2011uv,RindlerDaller:2009er,Li:2013nal}.

In this paper we examine whether it is plausible that the axion's correlation length grows dramatically. For definiteness, we will focus on the case in which the Peccei-Quinn phase transition happens after inflation, although the opposite ordering is also possible. We show that while long-range correlations can be established, in principle, for {\em repulsive} interactions, they do not occur for {\em attractive} interactions. Hence, although a Bose-Einstein condensate is still formed, a long-range order is not established. Our analysis applies to the QCD axion, but also applies to any bosonic dark matter particle whose behavior is dominated by attractive interactions. We demonstrate why the properties of the condensate are captured by classical field theory and we examine its equilibrium behavior. 

This paper is organized as follows: 
In Section \ref{Non-Rel} we introduce the non-relativistic field theory of axions. 
In Section \ref{Classical} we explain why the classical field approximation is valid. 
In Section \ref{Modes} we discuss the evolution of modes around a homogeneous background. 
In Section \ref{Ground} we discuss the equilibrium/ground state configurations.
In Section \ref{Realistic} we discuss the evolution from realistic initial conditions and provide coherence length estimates.
In Section \ref{Conclude} we summarize our results and discuss.
Finally, in the Appendix we include details of Friedmann-Robertson-Walker (FRW) expansion.

\section{Non-Relativistic Field Theory}\label{Non-Rel}

The axion is a scalar field $\phi$ introduced to solve the strong CP problem. At first approximation it is a massless Goldstone boson associated with a spontaneously broken global symmetry, but picks up a small mass due to non-perturbative effects in QCD. This leads to the following potential
\beq
V(\phi)=\Lambda^4(1-\cos(\phi/f_a))
\eeq
Here $\Lambda \sim 0.1$\,GeV is associated with the QCD scale, and $f_a$ sets the symmetry breaking scale. It can be shown that the abundance of axion dark matter in the universe is determined by $f_a$ with value
\beq
\Omega_a\sim\left(f_a\over10^{11-12}\;\mbox{GeV}\right)^{7/6}
\eeq
where the uncertainty in this expression is due to complications involved in calculating non-perturbative QCD effects, including the temperature dependence of the axion mass.

For small field values $\phi\ll f_a$, it is sufficient to expand the potential as follows
\beq
V(\phi)={1\over 2}m^2\phi^2+{\lambda\over 4!}\phi^4+\ldots
\eeq
where $m=\Lambda^2/f_a$ and $\lambda=-\Lambda^4/f_a^4<0$. Using this, we have the following relativistic Lagrangian density
\beq
\mathcal{L}={1\over2}(\partial\phi)^2-{1\over 2}m^2\phi^2-{\lambda\over 4!}\phi^4
\label{RelativisticLagrangian}\eeq

It is very useful to treat the axions in a non-relativistic
approximation, which is extremely well-justified.  Axions
interact far too weakly to be thermalized in the early universe,
so their production is dominated by the misalignment mechanism:
i.e., when the axion field acquires a mass during the QCD phase
transition, the phase of the field is generally misaligned with
the potential energy minimum.  As suggested by causality, the
field $\phi$ is expected initially to vary by an $\mathcal{O}(1)$
amount from one Hubble patch to the next, which implies that the
typical initial wavenumber $k_i\sim H_{\rm QCD}$, the Hubble
parameter at the QCD phase transition. Numerically, $H_{\rm QCD}
\sim T_{\rm QCD}^2/\mpl$, where $T_{\rm QCD}$ is the temperature
of the QCD phase transition, $T_{\rm QCD} \sim 0.1$ GeV, and
$\mpl \equiv 1/\sqrt{8 \pi G} \approx 10^{18}$ GeV is the reduced
Planck mass, so $k_i \sim 10^{-11}$ eV.  The axion mass increases
during the phase transition toward its final value $m$, typically
$\mathcal{O}(10^{-5}$ eV), so the axions are highly
non-relativistic shortly after the QCD phase transition.  $k$
redshifts with the scale factor, so for example by the time of
matter-radiation equality, $t \sim 50,000$ years, the typical
wavenumber is reduced further by factor of $\mathcal{O}(10^8)$. 
During structure formation, the axions are accelerated to
galactic speeds of $\mathcal{O}(10^{-3})\, c$, but the
non-relativistic approximation continues to be very accurate.

An important feature of the non-relativistic field theory
approximation is that particle-number violating processes are
ignored.  This is highly accurate, since the self-coupling
$\lambda = -\Lambda^4/f_a^4$ is extremely small: for $\Lambda\sim
0.1\,$GeV (typical QCD scale) and $f_a\sim 10^{11}$\,GeV (typical
Peccei-Quinn scale), we have $\lambda\sim -10^{-48}$.  The only
scattering process with an amplitude that is first order in
$\lambda$ is the particle-number preserving process
$2\phi\to2\phi$, since $\phi\to3\,\phi$ and $3\,\phi\to\phi$ are
kinematically forbidden.  Particle-number changing processes,
such as the annihilation process $4\phi\to 2\phi$, have cross
sections that are suppressed by an extra factor of $\lambda^2$. 
When photon couplings are included, the relativistic axion can
decay to two photons, but the lifetime is estimated as $\tau \sim
(m/20\ \mathrm{eV})^5$ times the age of the universe
\cite{PDGaxions2014}, which is $10^{28}$ times the age of the
universe for a typical mass of $10^{-5}$ eV\null.  Thus, all
particle-number violating processes can be very safely ignored.

In order to take the non-relativistic limit, let's re-write the real field $\phi$ in terms of a complex field $\psi$ as follows
\beq
\phi({\bf x},t)={1\over\sqrt{2m}}\left(e^{-imt}\psi({\bf x},t)+e^{imt}\psi^*({\bf x},t)\right)
\eeq
We substitute this into eq.~(\ref{RelativisticLagrangian}) and dispense with terms that go as powers of $e^{-imt}$ and $e^{imt}$, as they are rapidly varying and average out to approximately zero.  We then obtain the following non-relativistic Lagrangian for $\psi$ 
\beq
\mathcal{L}={i\over2}(\dot\psi\psi^*-\psi\dot\psi^*)-{1\over2m}\nabla\psi^*\!\cdot\!\nabla\psi-{\lambda\over 16m^2}(\psi^*\psi)^2
\eeq
For these non-relativistic fields, the momentum conjugate to $\psi$ is $\pi=i\,\psi^*$.
Note that this Lagrangian only involves a single time derivative on the complex field $\psi$.

Passing to the Hamiltonian and promoting the physical quantities to operators for the purpose of quantization, we obtain
\beq 
\hat H=\hat H_{\rm kin}+\hat H_{\rm int}
\eeq 
where
\bea 
&& \hat H_{\rm kin}=\int\! d^3x{1\over 2m}\nabla\hat\psi^\dagger\!\cdot\!\nabla\hat\psi
\label{kinetic}
\\
&&\hat H_{\rm int}=\int\! d^3x{\lambda\over16m^2}\hat\psi^\dagger\hat\psi^\dagger\hat\psi\hat\psi
\label{potential}
\eea 
The first term represents kinetic energy and the second term represents a short range interaction; attractive for $\lambda<0$ and repulsive for $\lambda>0$. 

The local number density of particles is given by
\beq
\hat n({\bf x})=\hat\psi^\dagger({\bf x})\hat\psi({\bf x})
\eeq
and the corresponding mass density is $\hat\rho(x)=m\,\hat\psi^\dagger(x)\hat\psi(x)$. With this understanding, it is straightforward to guess the form of the gravitational contribution to the energy
\beq
\hat H_{\rm grav}=-{Gm^2\over2}\!\int\! d^3x\!\int\! d^3x'{\hat\psi^\dagger({\bf x})\hat\psi^\dagger({\bf x}')\hat\psi({\bf x})\hat\psi({\bf x}')\over|{\bf x}-{\bf x}'|}
\eeq
The total Hamiltonian is the sum
\beq 
\hat H=\hat H_{\rm kin}+\hat H_{\rm int}+\hat H_{\rm grav}
\label{FullHamiltonian}
\eeq 

Although we derived this Hamiltonian starting with fields, we can also derive it using the more fundamental starting point of many-particle quantum mechanics. Consider the following Hamiltonian for $N$ non-relativistic particles, interacting via a contact interaction and gravity
\beq
\hat H=\sum_{i=1}^N{\hat p_i^2\over 2m}+{\lambda\over8m^2}\sum_{i<j}\delta^3(\hat x_i-\hat x_j)-\sum_{i<j}{Gm^2\over|\hat x_i-\hat x_j|}
\eeq
It is useful to introduce creation and annihilation operators that act on particle states in the usual way and satisfy standard commutation relations
\beq
[\hat a_k,\hat a^\dagger_{k'}]=(2\pi)^3\delta^3(k-k')
\eeq
Later we will make use of the following dimensionless occupancy number
\beq
\hat{\mathcal{N}}_k=\hat a_k^\dagger\, \hat a_k/V
\eeq
where $V$ is the volume of the box in which the field theory lives.
The kinetic energy can be written in an obvious way
\beq 
\hat H_{\rm kin}=\int\!{d^3 k\over(2\pi)^3}{k^2\over 2m} \,\hat a_k^\dagger \, \hat a_k
\eeq 
and there is a similar representation for the other terms.

We can then pass to the field language by defining
\beq
\hat\psi(x)\equiv\int\! {d^3 k\over (2\pi)^3}\, \hat a_k \, e^{i kx}
\eeq
and obtain the field representation of the Hamiltonian in equation~(\ref{FullHamiltonian}).

\section{Classical Field Theory Approximation}\label{Classical}

Let us decompose the quantum field $\hat\psi$ as
\beq
\hat\psi=\psi+\delta\hat\psi
\eeq
where $\psi$ is the expectation value in a given state $\langle\hat\psi\rangle=\psi$ and $\delta\hat\psi$ is the quantum correction.
We would like to estimate the relative size of the quantum correction to the classical piece.

\subsection{Occupancy Number}

For coherent states with occupancy number $\mathcal{N}$ the typical relative size of the quantum correction for modes on scales of the typical wavelengths is
\beq
{\delta\hat\psi\over\psi}\sim{1\over\sqrt{\mathcal{N}}}
\eeq
This relative quantum correction has an interpretation as an analogue of ``shot-noise" that occurs for photon fluctuations around the classical electromagnetic field.

Hence we would like to estimate the occupancy number. For axions in our galaxy, the number density is given by
\beq
n_{\rm gal}={\rho_{\rm gal}\over m}\approx {\text{GeV}/\text{cm}^3\over 10^{-5}\,\text{eV}}={10^{14}\over \text{cm}^3}
\eeq
For typical viralized particles in the galaxy, the de Broglie wavelength is given by
\beq 
\lambda_{dB}={2\pi\over m v}\approx{2\pi\over 10^{-5}\,{\rm eV}\times 10^{-3}}\approx 10^4\,{\rm cm}
\eeq 
The characteristic occupancy number is then given by
\beq 
\mathcal{N}\sim n_{\rm gal}\lambda_{dB}^3\approx 10^{26}
\eeq 
which is huge. This says that in the galaxy today, axions are in the high occupancy number regime.
In fact in the early universe, before galaxy formation, the typical occupancy number was even higher, since the typical axion velocity was lower, which enhances the de Broglie wavelength; we shall discuss this in Section \ref{Coherence}.

In this very high occupancy regime, the relative size of the quantum corrections are very small. This means we should be able to just use the classical field theory. So let's return to the field representation of Section \ref{Non-Rel} and drop the ``hats" on $\psi$. Then using the Hamilton-Jacobi equations, we obtain the following approximate equation of motion
\beq
i\,\dot\psi=-{1\over 2m}\nabla^2\psi+{\lambda\over 8m^2}|\psi|^2\psi-Gm^2\psi\!\int\! d^3x'{|\psi({\bf x}')|^2\over|{\bf x}-{\bf x}'|}
\eeq
This is rather more complicated than the standard one-particle Schr\"odinger equation; this equation is non-linear and non-local.

\subsection{Free Theory Thermalization}\label{Thermal}

Having turned to the classical field theory, one might be concerned that it misses essential aspects of thermalization. Indeed, one might be concerned that one cannot see the details of any phase transition to a BEC. Strictly speaking, ordinary classical fields do not thermalize due to the Rayleigh-Jeans catastrophe at high wave numbers. 

However, if we cut off the theory at some high wavenumber $\kuv$, there is normally a well defined thermal equilibrium. In fact the classical  theory is able to describe the phase transition. To see this, let us consider a free field theory in contact with an external heat bath at temperature $T$.
The free energy functional is
\beq
F[\psi] = \int\! {d^3k\over(2\pi)^3}\left[{k^2\over 2m}-\mu(T)\right]|\psi_k|^2
\eeq
where $\mu(T)$ is the chemical potential.
The expectation of the number of particles is given by the ratio of path integrals
\beq
\langle N\rangle={\int \mathcal{D}\psi\, N[\psi]\, \exp\left(-F[\psi]/T\right)\over\int \mathcal{D}\psi\, \exp\left(-F[\psi]/T\right)}
\eeq
where the number functional is
\beq
N[\psi]= \int\! {d^3k\over(2\pi)^3} |\psi_k|^2
\eeq
Carrying out the path integrals, and dividing by a volume factor, we obtain the number density $n_{\rm th}$ of thermal particles
\beq 
n_{\rm th} = \int\! {d^3k\over(2\pi)^3}{T\over {k^2\over 2m}-\mu(T)}
\eeq 
Cutting off the integral at $|{\bf k}|=\kuv$ we obtain
\beq 
n_{\rm th} = {m \,T \,\kuv\over\pi^2}\left[1-{\sqrt{2m|\mu(T)|}\over \kuv}\tan^{-1}\left(\kuv\over\sqrt{2m|\mu(T)|}\right)\right]
\eeq 
with $\mu(T)\leq0$. So long as the total number density of particles is $n_{\rm tot}<mT\kuv/\pi^2$, we can always solve this equation for $\mu(T)$, implying that all particles are thermal. However if $n_{\rm tot}>mT\kuv/\pi^2$, then $\mu(T)$ is stuck at $\mu=0$ and not all particles can be thermal; there must be a {\em condensate} of particles in the ground state. In the free theory, the ground state is the $k=0$ mode. (Later we discuss the radical change that occurs when attractive interactions are included). The critical temperature for the phase transition is evidently
\beq 
T_{\rm crit} = {\pi^2\, n_{\rm tot}\over m\,\kuv}
\eeq 
The classical theory does not determine the cutoff $\kuv$, but if we adopt an estimate from quantum theory, $\kuv^2/2m = T_{\rm crit}$, then we find $T_{\rm crit}=(\pi^4/2)^{1/3} n_{\rm tot}^{2/3}/m$, which differs by only 10\% from the quantum mechanical answer, $T_{\rm crit}= 2 \pi \bigl(n_{\rm tot}/\zeta(3/2)\bigr)^{2/3}/m$.
If we keep the density of particles and the cutoff fixed, then the ratio of the number of particles in the ground state condensate $n_c$ to the total number of particles $n_{\rm tot}$ is linear in the temperature $T$, and given by
\beq 
{n_c(T)\over n_{\rm tot}}=
\Bigg{\{}\begin{array}{c}
\,\,\,\,\,\,\,0\,\,\,\,\,\,\,\,\,\,\,\,\,\,\,\mbox{for}\,\,\,T>T_{\rm crit}\\
1-{T\over T_{\rm crit}}\,\,\,\mbox{for}\,\,\,T<T_{\rm crit}
\end{array}
\eeq 
So for $T\ll T_c$ almost all particles are in the condensate. Since cosmological axions are at very high density and are non-relativistic, we expect $ T\ll T_{\rm crit}$, if indeed thermal equilibrium is established.

The two-point correlation function $\langle \psi^*({\bf x})\,\psi({\bf y})\rangle$ can also be computed in terms of the chemical potential. In integral form it is
\beq
\langle \psi^*({\bf x})\,\psi({\bf y})\rangle = \int\! {d^3k\over(2\pi)^3}{T\over {k^2\over 2m}-\mu(T)}e^{i{\bf k}\cdot({\bf x}-{\bf y})}+n_c(T)
\eeq
where we have separated out the thermal piece and the condensate piece.
In the short distance limit $|{\bf x}-{\bf y}|\to0$ this is just the number density, as we computed above, and is sensitive to the value $\kuv$. On the other hand, in the long distance limit $|{\bf x}-{\bf y}|\gg 1/\kuv$, the dependence on $\kuv$ is less important and only appears implicitly through $\mu(T)$. In this limit the two-point correlation function is
\beq
\langle \psi^*({\bf x})\,\psi({\bf y})\rangle = {mT\over2\pi|{\bf x}-{\bf y}|}e^{-\sqrt{2m|\mu(T)|}|{\bf x}-{\bf y}|}+n_c(T)
\eeq
So for $ T>T_{\rm crit}$, with $|\mu(T)|>0$ and $n_c(T)=0$, the correlation function falls off exponentially with distance and has a finite correlation length. For $ T=T_{\rm crit}$, with $ \mu(T_{\rm crit})=0$ and $ n_c(T_{\rm crit})=0$, the correlation function falls off as a power law. 
For $ T<T_{\rm crit}$, with $\mu(T)=0$ and $n_c(T)>0$, the correlation function asymptotes to a non-zero value at large distances. Hence there is tremendous long-range correlation for $ T<T_{\rm crit}$. In this paper we shall examine how this is altered in the interacting theory and how it depends on the sign of the interaction.

In summary, the classical field theory can adequately describe the phase transition from a regular phase to a BEC. While a BEC is a very quantum phenomenon from the {\em particle} point of view, it is a very classical phenomenon from the {\em field} point of view. 
By including interactions, we should be able to understand the formation of the BEC, or otherwise, and its properties, purely by studying the classical field theory.

\section{Evolution Around Homogeneous Condensate}\label{Modes}

In some of the simplest and most familiar BECs, such as those described by a free theory coupled to an external heat bath,
the system is driven to an equilibrium state where almost all particles are in the $k=0$, or very small $k$, modes. From the classical field point of view, this means that the field $\psi$ is driven to be very slowly varying in space, with extremely long-range correlation.
If the system is entropically driven to such an equilibrium configuration, it must be stable against perturbations. In this section, we examine whether this applies to the axion using linear perturbation theory.

\subsection{Self-Interactions}

Let us begin by considering the contact interaction and ignore gravity. The equation of motion for the classical field is
\beq
i\,\dot\psi=-{1\over 2m}\nabla^2\psi+{\lambda\over 8m^2}|\psi|^2\psi
\label{NLS}\eeq

Let's decompose the field into a homogeneous piece $\psi_c$ and a perturbation $\delta\psi$ as
\beq
\psi({\bf x},t)=\psi_c(t)+\delta\psi({\bf x},t)
\label{expansion}\eeq
The homogeneous piece is, effectively, the condensate, while $\delta\psi$ represents a small disturbance in it.
The condensate satisfies the equation
\beq 
i\,\dot\psi_c={\lambda\over 8m^2}|\psi_0|^2\psi_c
\eeq 
This has a simple periodic solution
\beq
\psi_c(t)=\psi_0\, e^{-i\,\mu_c\, t}
\eeq
where 
\beq 
\mu_c={\lambda\over 8m^2}|\psi_0|^2
\eeq 
The prefactor $\psi_0$ is not a free parameter; its magnitude is determined by $n_0=|\psi_0|^2$, where $n_0$ is the density of particles. On the other hand, the phase of $\psi_0$ is arbitrary. Any choice for the phase spontaneously breaks the global $U(1)$ symmetry associated with particle number conservation.

Perturbing the differential equation (\ref{NLS}) to linear order leads to
\beq
i\,\dot{\dPsi}=-{1\over 2m}\nabla^2\dPsi+{\lambda\,n_0\over 8\,m^2}(\dPsi+\dPsi^*)
\label{linear}\eeq
where, for convenience, we have traded $\delta\psi$ for $\delta\Psi$ through $\delta\psi=\psi_c\,\dPsi$.
Now we decompose $\dPsi$ into real and imaginary parts as
\beq
\dPsi=A+iB
\eeq
Then after Fourier transforming, we obtain
\beq 
{d\over dt}\left(\begin{array}{c}A_k\\B_k\end{array}\right)
=\left(\begin{array}{cc}0 & {k^2\over 2m}\\-{k^2\over 2m}-{\lambda n_0 \over 4m^2} & 0\end{array}\right)\left(\begin{array}{c}A_k\\B_k\end{array}\right)
\eeq 
Depending on the sign of 
\beq
\kappa_k \equiv \frac{k^2}{2m} + \frac{\lambda n_0}{4 m^2} 
\eeq
the solutions have one of two possible forms.  For $\kappa_k < 0$, the solutions are pure exponentials,
\beq
\dPsi_k=c_1 (\gamma_k-i \kappa_k) \,e^{\gam_k t}+c_2 (\gamma_k+i \kappa_k)\,e^{-\gam_k t }
\label{pure-exponential}
\eeq
where $c_1$ and $c_2$ are arbitrary real constants, 
$\pm \gam_k$ are the eigenvalues of the above matrix,
\beq
\gam_k=\frac{k}{\sqrt{2m}}\sqrt{-\kappa_k}
\label{Floquet}
\eeq
and $(\gamma_k \mp i \kappa_k)$ are the eigenvectors.
For $\kappa_k > 0$ we can begin with a trial function of the form $\dPsi_k = Z_1 e^{- i \omega_k t} + Z_2 e^{i \omega_k t}$, which leads to the solution
\beq
\dPsi_k = Z (\omega_k + \kappa_k) e^{- i \omega_k t} +  Z^* (\omega_k - \kappa_k) e^{i \omega_k t}
\label{oscillation}
\eeq
where
$Z$ is an arbitrary complex constant, and
\beq
\omega_k = \frac{k}{\sqrt{2m}} \sqrt{\kappa_k}
\eeq

Hence, if $\lambda < 0$, the modes for $k$ values in the range
\beq
k^2<-{\lambda\,n_0\over 2\,m}
\eeq
 experience parametric resonance and there is exponential growth of perturbations. For higher values of $k^2$ there is no growth, only oscillations in the perturbation.

The existence of an instability band is therefore determined by the sign of the self-coupling $\lambda$. In summary we have
\bea
&&\lambda>0\implies\mbox{stability}\\
&&\lambda<0\implies\mbox{instability}
\eea
For the QCD axion, we have $\lambda<0$. This is an {\em attractive} interaction, and hence there is an instability.
This means the homogeneous condensate with long-range correlation is {\em not} an attractor configuration of the system. It is therefore not the entropically preferred configuration that arises dynamically through thermalization.
Similar remarks go through for very small, but non-zero, $k$ modes. On the other hand for systems with $\lambda>0$, the homogeneous configuration is stable and is an attractor solution under thermalization.

\subsection{Gravity}

We now investigate the case of  gravity, and ignore the self-coupling $\lambda$.
The equations we need to solve are
\bea
i\,\dot\psi &= & -{1\over 2m}\nabla^2\psi+m\,\phi_N\psi\\
\nabla^2\phi_N &=& 4\pi G (m |\psi|^2-\bar{\rho})
\eea
where we have subtracted out the average background density $\bar\rho$ in the equation for the Newtonian potential, which will be appropriate in the FRW analysis given in Appendix~\ref{Bgd}.

Expanding $\psi$ as before in eq.~(\ref{expansion}) we have the trivial solution for the condensate
\beq
\psi_c(t)=\psi_0\,\,\,(\mbox{constant})
\eeq
The linearized equations for the fluctuations are
\bea
i\,\dot\dPsi &= & -{1\over 2m}\nabla^2\dPsi+m\,\phi_N\\
\nabla^2\phi_N &=& 4\pi G\, m \,n_0(\dPsi+\dPsi^*)
\eea
Eliminating $\phi_N$ leads to 
\beq
i\,\dot\dPsi=-{1\over 2m}\nabla^2\dPsi+4\pi G m^2 n_0\nabla^{-2}(\dPsi+\dPsi^*)
\eeq
which is identical to the structure of eq.~(\ref{linear}) in the $\lambda\,\phi^4$ theory with the replacement
\beq
{\lambda\over 8 m^2}\leftrightarrow 4\pi G m^2\nabla^{-2}
\eeq
By Fourier transforming, and using the result in eq.~(\ref{Floquet}), we obtain
\beq
\gam_k={k\over 2m}\sqrt{{16\pi G m^3n_0\over k^2}-k^2}
\label{FloquetGrav}\eeq
So again we find instability for a condensate of long-range correlation. Modes that satisfy
\beq
k<k_J = (16\pi G m^3n_0)^{1/4}
\label{Jeans}
\eeq
are unstable. Here $k_J$ is a type of Jeans wavenumber, as it separates the regime where gravity dominates, leading to collapse, and the regime where pressure dominates, leading to oscillations.  This pressure is, from the particle point of view, a type of 
``quantum pressure," arising from the uncertainty principle: even though the background particles are at rest, a perturbation of wavelength $2\pi/k$ implies that at least some of the particles are localized on this distance scale, requiring an increase in the energy, with the accompanying restoring force.

We note that if we were to send $G\to-G$, and consider repulsive gravity, then the condensate would be stable. It would in fact be an attractor solution; entropically favored under thermalization. Although repulsive Newtonian gravity is unphysical, we know that in general relativity, we can achieve effective repulsion provided by vacuum energy, as is the case during inflation \cite{Guth:1980zm,Linde:1981mu,Albrecht:1982wi}. In this case, the field organizes into a type of condensate with tremendously long-range correlation. (There have also been interesting examples of this with light vector fields \cite{Nelson:2011sf}).

\subsection{Occupancy Number Evolution}

We can gain further understanding of the behavior of a perturbed condensate by tracking the evolution of the occupancy number 
\beq
\mathcal{N}_k=|\psi_k|^2/V
\eeq
for each mode.  We can use the linearized evolution of equations~(\ref{pure-exponential}) and (\ref{oscillation}), choosing an initial perturbation (at time $t_i$) with a random phase.  Writing $\delta \Psi_k(t_i) \equiv A_k + i B_k$, a randomized phase implies that $\langle A_k^2 \rangle = \langle B_k^2 \rangle \equiv \sigma^2_k/2$, and $ \langle A_k B_k \rangle = 0$, which with equation~(\ref{pure-exponential}) implies that $\langle c_1^2 \rangle = \langle c_2^2 \rangle = \sigma_k^2 (\kappa_k^2+ \gamma_k^2) / 8 \gamma_k^2 \kappa_k^2$ and $\langle c_1 c_2 \rangle = \sigma_k^2 (\kappa_k^2 - \gamma_k^2) / 8 \gamma_k^2 \kappa_k^2$.  We then find that for $\kappa_k < 0$ the occupancy number evolves (for $k\neq 0$) as
\beq
\langle\mathcal{N}_k(t)\rangle = \langle\mathcal{N}_k^i\rangle\left[1+{1\over2}\left(\lambda\, n_0\over4\,m^2\right)^{\!2}{\sinh^2(\gam_k\, (t-t_i))\over\gam_k^2}\right]
\label{OccupancyLinear}\eeq
where $\langle\mathcal{N}_k^i\rangle$ is the initial value of $\langle\mathcal{N}_k(t)\rangle$ at $t=t_i$.  For $\kappa_k > 0$ we use equation~(\ref{oscillation}) with $\langle Z^2 \rangle = \langle Z^{*2} \rangle = \sigma_k^2 (\kappa_k^2 - \omega_k^2)/8 \omega_k^2 \kappa_k^2 $ and $\langle Z Z^* \rangle = \sigma_k^2 (\kappa_k^2 + \omega_k^2 )/8 \omega_k^2 \kappa_k^2$, finding the same result, provided that we use $\gamma_k = i \omega_k$, so $\sinh^2 (\gamma_k (t-t_i))/\gamma_k^2 = \sin^2 (\omega_k (t-t_i))/ \omega_k^2$.
So at early times, $|\gam_k|\,(t-t_i)\ll1$, the sign of $\lambda$ is unimportant and the occupancy numbers grow as $\sim (t-t_i)^2$. However at late times, $|\gam_k|\,(t-t_i)\gg1$, there is oscillatory or exponential behavior depending on the sign of $\lambda$ and the $k$-mode. 
\begin{itemize}
\item For $\lambda>0$, $\gam_k$ is imaginary for all $k$ and $\sinh^2(\gam_k (t-t_i))/\gam_k^2\to \sin^2(\omega_k(t-t_i))/\omega_k^2$, so the occupancy number undergoes stable oscillations.  Since we have averaged over phases, it may seem surprising that we see net growth starting from $t=t_i$; indeed if we had randomized the phase of $Z$ instead of $\dPsi$, we would have found a time-independent occupancy number.  The phases are related in such a way that a random phase for $\dPsi$ results in the phase of $Z$ being more likely to be at the low end of the occupancy number oscillations.  If we had considered {\it any} specific solution, without averaging over phases, we would have seen larger oscillations: from equation~(\ref{oscillation}), one can show that
\beq
{\langle \mathcal{N}_k(t) \rangle_{\rm max} \over {\langle
\mathcal{N}_k(t) \rangle_{\rm min}}} = 1 + {1 \over \omega_k^2}
\left[{\lambda n_0 \over 4 m^2} \left( {k^2 \over 2m} + {\lambda
n_0 \over 4 m^2} \right)\right]
\eeq
which means that the oscillations for any solution are at least twice as large as the phase-averaged oscillations shown in equation~(\ref{OccupancyLinear}).  The important point, however, is that the oscillations are stable.
The largest ratio of $\langle\mathcal{N}_k(t)\rangle/\langle\mathcal{N}_k^i\rangle$ is obtained for the modes that minimize $\omega_k$, which occurs as $k\to 0$, as the amplitude scales as $\sim 1/k^2$. Hence low $k$-modes dominate and the homogeneous condensate, or more generally the configuration dominated by long-range correlations, is stable.
\item For $\lambda<0$, $\gam_k$ is real for a band of $k$ and $\sinh(\gam_k(t-t_i))$ grows exponentially for these modes. Hence the fastest growth is for the modes that {\em maximize} $\gam_k$, which occurs at $k=k_*$, where $k_*$ is given below as equation~(\ref{kstar}.) Hence these finite $k$-modes dominate and cause the system to evolve towards localized clumps, as we describe in the next section.
\end{itemize}

Similar statements go through for  gravity.

\section{Ground States}\label{Ground}

When the couplings are attractive, the equilibrium/ground state of the system is not a homogeneous condensate but a localized clump \cite{Khlebnikov:1999qy,Khlebnikov:1999pt,Chavanis:2011zi,Chavanis:2011zm}. Its structure is different for the case of self-interactions and  gravity, as we now describe.

\subsection{Solitons}

For bosons with self-coupling $\lambda<0$ the system is unstable toward fragmenting into a complicated configuration governed by a range of wave numbers. The growth rates are maximized at 
\beq
k_*=\sqrt{|\lambda|\,n_0\over4\,m}
\label{kstar}
\eeq
This sets the characteristic scale at which structures should form. 

In 1+1 dimensions this can lead to the production of stable solitons: ground state configurations at fixed number of particles. 
For a soliton $\psi_s$ comprised of $N$ particles, the solution in its center-of-mass frame is
\beq
\psi_s(x,t) = \sqrt{k_s\,N\over2}\,\mbox{sech}(k_s\, x)\,e^{-i\,\mu_s\, t}
\eeq
where 
\bea 
k_s &=& {|\lambda|\,N\over 16\,m}\label{ks}\\
\mu_s &=& - {k_s^2\over 2 m}
\eea 
and the ground state energy, as defined by equations~(\ref{kinetic}) and (\ref{potential}), is
\beq
E_s = -{\lambda^2N^3\over 1536m^3}
\eeq
This solution is known as a ``Bright soliton". The wavenumber $k_*$, associated with maximal growth away from the homogeneous configuration, is of the same order as the dominant wavenumber that comprises the soliton $k_s$. To see this, note that the  core of the soliton has characteristic number density $n_s\sim k_s\,N$. If we re-arrange this as $N\sim n_s/k_s$, insert into equation (\ref{ks}), and solve for $k_s$, we find that parametrically $k_s\sim k_*$.

We note that BEC's do not usually form in 1+1 dimensions. In fact if one returns to the free theory analysis of Section \ref{Thermal} and repeats the analysis in 1 spatial dimension, one finds no actual phase transition. More interesting is to go to 3+1 dimensions, where a phase transition can take place. But then the solitons are not exactly stable. Without further refinement, they are subject to a collapse instability. In the case of the axion, one can produce so-called ``axitons" in the early universe \cite{Kolb:1993hw}, which have finite lifetime.

As we describe in Section \ref{Relax}, in axion cosmology the claim of thermalization to a BEC comes from considerations of gravitational interactions, to which we now turn.

\subsection{Bose stars}

For ordinary (attractive) gravity the system  tends to fragment, in an analogous way to the case with self-coupling.
In this case it can lead to a stable bound state in 3-dimensions held together by gravity: a ``Bose star". The Hamiltonian for these gravitationally bound configurations $\psi_g$ is
\beq
H=\int \!d^3x {|\nabla\psi_g|^2\over 2m}-{Gm^2\over2}\!\int\!d^3x\!\int\! d^3x'{|\psi_g({\bf x})|^2|\psi_g({\bf x}')|^2\over|{\bf x}-{\bf x}'|}
\eeq
The ground state comes from minimizing the Hamiltonian at fixed particle number $N$.
We do not know an exact solution for this system of equations. However, a variational approximation will suffice.
The ground state will be spherically symmetric $\psi({\bf x})=\psi(r)$. As a variational ansatz, we take its profile to be exponential (mimicking the ground state wavefunction of the hydrogen atom)
\beq
\psi_g(r) = \sqrt{N\,k_g^3\over\pi}\,e^{-k_g\,r}e^{-i\,\mu_g\,t}
\eeq
where $k_g$ is a variational parameter that has units of wavenumber.
Substituting into the Hamiltonian and carrying out the integrals, we obtain
\beq
H = {Nk_g^2\over 2m}-{5Gm^2N^2k_g\over 16}
\eeq
Extremizing $H$ with respect to $k_g$, we obtain the characteristic wavenumber of the Bose star
\beq
k_g={5Gm^3N\over16}\label{kg}
\eeq
and the corresponding approximation for the ground state energy 
\beq
E_g = -{25G^2m^5N^3\over512}
\eeq
As in the case of the soliton, the characteristic wavenumber $k_g$ of the ground state $\psi_g$ is connected to the characteristic wavenumber $k_J$ of the exponentially growing modes away from the homogeneous condensate $\psi_c$. To see this, note that in the core of the Bose star, the number density $n_g$ satisfies $N\sim n_g/k_g^3$; inserting this into equation (\ref{kg}) and solving for $k_g$, we have $k_g\sim k_J$.

\subsection{Characteristic Wavenumber Summary}

\begin{figure}[t]
\includegraphics[width=\cw]{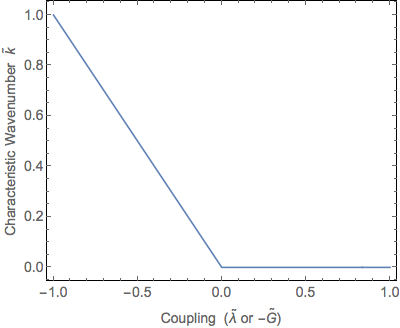}
\caption{The characteristic wavenumber $\tilde k\equiv k/m$ of the ground state as a function of the coupling; self-interactions $\tilde\lambda$ (in 1+1 dimensions $\tilde\lambda$ is normalized as $\tilde\lambda\equiv\lambda N/(16m^2)$) or  gravity $-\tilde G$ (in 3+1 dimensions $\tilde G$ is normalized as $\tilde G\equiv 5Gm^3N/(16m)$). For repulsive interactions ($\lambda>0$ or $G<0$) the ground state is governed by $k=0$. For attractive interactions ($\lambda<0$ or $G>0$) the characteristic wavenumber is non-zero and given in equations (\ref{ks},\,\ref{kg}). } 
\label{CharacteristicWavenumber}\end{figure}

A summary of the dependence of the typical wavenumber of the ground/equilibrium state is given in Figure \ref{CharacteristicWavenumber}. For repulsive interactions, the ground state is governed by $k=0$, while for attractive interactions the ground state is governed by wavenumbers given in equations (\ref{ks},\,\ref{kg}). We note that for attractive, but very small couplings, the ground state is still very homogeneous, governed by large, but not infinite, wavelengths. For large couplings, the ground states are rather compact. We shall estimate the relevant scale for the axion in Section \ref{Coherence}, and explain why these characteristic wavelengths (inverse wavenumber) also set the typical correlation length.

\section{Evolution for Realistic States}\label{Realistic}

The previous analysis shows that a condensate with long-range correlation is {\em not} the attractor point in phase space for the axion. Instead the attractor point in phase space includes Bose clumps: solitons or stars. 
In this section we investigate the behavior starting from some plausible initial conditions. 

The axion is a Goldstone boson that arises after the Peccei-Quinn symmetry is broken. Assuming this happens after inflation, we expect the axion field to be initially distributed randomly from one Hubble patch to the next, as causality forbids any initial super-horizon correlations. (While inflation allows the possibility of super-Hubble correlations, we assume that inflationary-era correlations have no significant influence on the order that arises in the post-inflationary Peccei-Quinn phase transition.)  In a given Hubble patch, the axion field should be fairly uniform as gradients are energetically disfavored. 
This suggests a form of {\em white noise} initial conditions with a UV cutoff $k_{\mbox{\tiny{UV}}}\sim H_i$, where $H_i$ is the Hubble parameter at the time of formation. 

For simplicity, we assume the axion is initially drawn from a Gaussian distribution. It has a non-zero two-point function given by
\beq
\langle\psi({\bf k},t)\,\psi^*({\bf k}',t)\rangle = (2\pi)^3\delta^3({\bf k}-{\bf k}')\langle\mathcal{N}_k(t)\rangle
\eeq
Here $\langle\mathcal{N}_k(t)\rangle)$ is usually called the power spectrum $P(k,t)$.
We also assume that initially ($t=t_i$) the real and imaginary parts of $\psi$ are uncorrelated and identically distributed, meaning that the autocorrelation function is trivial
\beq
\langle\psi({\bf k},t_i)\,\psi({\bf k}',t_i)\rangle =0
\eeq
At later times, $t>t_i$, the real and imaginary parts can become correlated, and the autocorrelation function can become non-zero.
The specific form of the initial power spectrum $\langle\mathcal{N}_k^i\rangle$ is not important for our discussion, but a reasonable choice would be the following
\beq 
\langle\mathcal{N}^i_k\rangle = {(2\pi)^{3/2}\,n_{\rm ave} \over k_{\mbox{\tiny{UV}}}^3}\exp\left(-k^2/(2\,k_{\mbox{\tiny{UV}}}^2)\right)
\eeq 
where $n_{\rm ave}$ is the average density of particles. For $k\ll\kuv$ the spectrum is flat, which is white noise.
As long as the prefactor $n_{\rm ave}/k_{\mbox{\tiny{UV}}}^3\gg1$ then the occupancy of modes with $k<k_{\mbox{\tiny{UV}}}$ is large and the classical field theory is adequate to describe these modes.

\subsection{Relaxation Rate}\label{Relax}

Since the white noise initial distribution for the axion is rather incoherent on large scales, the evolution of modes is more complicated than that of the previous section. However the previous analysis contains some of the central information in it, as we now explain.

For the case of self-interaction, the equation governing the evolution of modes is
\beq
i\,\dot\psi_k = {k^2\over 2m}\psi_k+{\lambda\over 8m^2}\!\int\!{d^3k'\over(2\pi)^3}\!\int\!{d^3k''\over(2\pi)^3}\,\psi_{k'}\psi_{k''}^*\psi_{k+k''-k'}
\label{timeev}\eeq
The evolution of the occupancy number $\mathcal{N}_k=|\psi_k|^2/V$ is then given by
\beq
\dot{\mathcal{N}}_k = -{\lambda V^{-1}\over8m^2}\!\!\int\!{d^3k'\over(2\pi)^3}\!\int\!{d^3k''\over(2\pi)^3}\left[i\psi_{k'}\psi_{k''}^*\psi_{k+k''-k'}\psi_k^*+c.c\right]
\label{occev}\eeq
Drawing $\psi_k$ from an initially Gaussian distribution, with initially independent real and imaginary parts, we find the expectation value of the 1st time derivative is initially zero
\beq
\langle\dot{\mathcal{N}}^i_k\rangle = 0
\eeq

However the expectation value of the 2nd time derivative is initially non-zero. By taking a time derivative of equation (\ref{occev}), then using equation (\ref{timeev}), then taking an expectation value and using Wick's theorem, we find it to be
\bea 
\langle\ddot{\mathcal{N}}^i_k\rangle &=& \left(\lambda\over4\,m^2\right)^{\!2}\Bigg{[}-
n_{\rm ave}^2\,
\langle\mathcal{N}_k^i\rangle\nonumber\\
&&+\int\!{d^3k'\over(2\pi)^3}\!\int\!{d^3k''\over(2\pi)^3}
\langle\mathcal{N}_{k'}^i\rangle\langle\mathcal{N}_{k''}^i\rangle\langle\mathcal{N}_{k-k'-k''}^i\rangle\Bigg{]}\,\,\,\,\,\,
\label{2ndDeriv}
\eea 
(we have used $n_{\rm ave}=\int\!{d^3k'\over(2\pi)^3}\mathcal{N}_{k'}$ to simplify the first term.)
This allows us to estimate a kind of ``relaxation rate": the typical rate at which modes are initially changing. 
By estimating $\Gamma_k\sim\sqrt{|\langle\ddot{\mathcal{N}}^i_k\rangle/\langle\mathcal{N}^i_k\rangle|}$, we find that a typical value for the bulk of the modes is
\beq 
\Gamma_k \sim {|\lambda|\,n_{\rm ave}\over 4\,m^2}
\eeq 
At this early time (and for this special choice of initial conditions), the evolution is independent of the sign of $\lambda$. But we know that the late time equilibrium behavior (homogeneous condensate or localized clump) is entirely controlled by the sign of $\lambda$, as we showed in the previous sections. 
Indeed the dependence on the sign of $\lambda$ can be seen by going to higher time derivatives.

Note that this relaxation rate $\Gamma_k$ is the same prefactor that appears in equation (\ref{OccupancyLinear}) for the evolution of modes around the homogeneous condensate, with $ n_0\to n_{\rm ave}$. 
Also, by replacing $\lambda/(8m^2)\to -4\pi Gm^2/k^2$ appropriately inside the convolution integrals, we obtain the gravitational case
\beq 
\Gamma_k\sim{8\pi G\,m^2 \,n_{\rm ave}\over k^2}
\eeq 
For the gravitational case, the rate is relatively large at late times because the wavenumber redshifts, so there is a relative enhancement of $a^2$ and it grows. This was noted in Refs.~\cite{Sikivie:2009qn,Erken:2011dz} and provided much of the motivation for the claims of thermalization. For this reason we will focus on this later in Section \ref{Coherence}.

\subsection{Thermalization}

The nonlinear evolution of the initial mess of white noise modes is presumably associated with some form of thermalization. Since the system is at high occupancy, the associated temperature is well below the critical temperature $T_c$, and so the system tries to organize into some form of BEC.

\begin{itemize}

\item For $\lambda>0$ (or repulsive gravity), the thermalization is towards a condensate with almost all particles in the ground state with $k=0$ (or very small $k$) -- a homogeneous configuration with long-range correlation. (For $\lambda>0$ this was nicely seen in the numerical work of Ref.~\cite{Berges:2012us}. But it was later, in Ref.~\cite{Berges:2014xea}, applied incorrectly to the axion, for which the interactions have the opposite sign.)

\item For $\lambda<0$ (or attractive gravity), there is a ``bottleneck" to achieve thermalization. In true equilibrium, the field would organize into a condensate with almost all particles in the ground state; this would be a single extremely compact clump of well-defined phase, as described in Section \ref{Ground} (equations (\ref{ks},\,\ref{kg}) show that the ground state has width that is inversely proportional to the number of particles $N$). However, as a coherent clump is forming locally in one region of space, its local equilibrium means that it stops re-organizing the phase in distant regions of space. So the phases of distant regions can remain uncorrelated. It is therefore difficult to achieve true global thermal equilibrium. Instead one expects only intermittent patches of coherent clumps (solitons or stars) made up of a moderate number of particles, along with a messy scalar field that has yet to reach true equilibrium. In any case, no long-range correlation is established. A full simulation of this process is ongoing work.

\end{itemize}

\subsection{Comparison to Boltzmann Equation}

We note that this critical dependence on the sign of $\lambda$ arises because we are in the classical field theory limit, which is applicable to dark matter axions. In other regimes, the sign of $\lambda$ can become relatively unimportant. 

For instance, in the particle limit, we can usually just track the classical {\em particle} phase space density $\mathcal{N}({\bf x},{\bf p})$, where one treats particles as carrying a well defined position and momentum (albeit perhaps allowing for a semi-classical enhancement from occupancy factors). The evolution of $\mathcal{N}({\bf x},{\bf p})$ is described by the Boltzmann equation, which governs the evolution to equilibrium. For non-relativistic $2\to2$ collisions, the evolution equation can be written as
\bea
{D\over Dt}\mathcal{N}_{p_1} \!&=\!& \int \!{d^3p_2\over(2\pi)^3}\, d\sigma\,v_{\rm rel} 
\Big{[}\mathcal{N}_{p_1'}\,\mathcal{N}_{p_2'}(1+\mathcal{N}_{p_1})(1+\mathcal{N}_{p_2}) \nonumber\,\,\,\,\,\,\,\,\,\,\, \\
&&\,\,\,\,\,\,\,\,\,\,\,\,\,\,\,\,\,\,\,\,\,\,\,\,\,\,\,\,\,\,\,\,\, - \mathcal{N}_{p_1} \mathcal{N}_{p_2}(1+\mathcal{N}_{p_1'})(1+\mathcal{N}_{p_2'})\Big{]}
\eea
where all the $\mathcal{N}_p$'s are evaluated at the same point in space ${\bf x}$. The typical rate of interaction is $ \Gamma\sim n_{\rm ave}\,\sigma\, v_{\rm rel}\,\mathcal{N}$. 

Since the particle scattering cross section $\sigma\propto\lambda^2$, the sign of $\lambda$ does not appear in this evolution equation. However, in the high density/coherent limit that is relevant for cosmological axions, we need to replace this semi-classical {\em particle} description with the classical {\em field} description. Then the sign of $\lambda$ plays a critical role in the evolution equation and dictates its equilibrium behavior, as discussed in the above sections.

\subsection{Coherence Length Estimate}\label{Coherence}

Having found that attractive interactions do not cause the system of axions to evolve to an equilibrium state of huge correlation length -- in contradiction to the conclusions of Refs.~\cite{Sikivie:2009qn,Erken:2011dz} -- we turn now to estimate the actual size of the correlation length for a universe comprised of radiation and axion-dominated matter, focussing on the physically relevant case of (attractive) gravity. A proper treatment would require a nonlinear simulation, but here we give a rough estimate of the length scales.

Initially the characteristic lengths evolve under the standard redshifting, so the physical wavenumbers scale as $k_{\rm phys}\sim k/a$. This continues until the relaxation rates become comparable to the Hubble expansion rate, $\Gamma_k\sim H$. At this point the system will attempt to thermalize, and form Bose stars -- although it is subject to the bottleneck described above.
An associated length scale for these Bose stars can be roughly estimated, as follows.

At the QCD phase transition, when the axion potential turns on, the characteristic wavenumber is $ k\sim H_{\rm QCD}\sim T_{\rm QCD}^2/\mpl$, where $\mpl \equiv 1/\sqrt{8 \pi G} \approx 10^{18}$ GeV is the reduced Planck mass. This also sets the initial correlation length. Assuming the axions comprised most of the matter in the universe, the number density of axions at this early time is $ n_a\sim\rho_a/m \sim (T_{\rm eq}/T_{\rm QCD})\rho_{\rm tot}/m\sim T_{\rm eq}T_{\rm QCD}^3/m$.  The number of axions within a typical de Broglie wavelength sets a typical occupancy number
\bea 
\mathcal{N}\sim {n_a\over k^3}\sim{T_{\rm eq}\mpl^3\over T_{\rm QCD}^3m}
&\sim&{0.1\,\mbox{eV}\times (10^{18}\,\mbox{GeV})^3\over (100\,\mbox{MeV})^3\times 10^{-5}\,\mbox{eV}}\,\,\,\,\\
&=&10^{61}
\eea 
At late times, once $\Gamma>H$, the system will attempt to thermalize. As we explained above, the bottleneck to thermalization means that only a fraction of the axions will organize into the ground state Bose stars. A rough estimate would be to take the occupancy number $\mathcal{N}$ as the typical number $N$ of axions that form  a Bose star, which is equivalent to saying that a typical Bose star contains the total energy of the axion field within one horizon volume at the time of the QCD phase transition. Furthermore, since there is no true equilibrium established and distant Bose stars maintain random phases, we expect the typical size of the Bose stars to roughly set the correlation length of the condensate.

Using equation (\ref{kg}) we see that the typical wavelength of such Bose stars, and hence the associated correlation length, is roughly
\bea
\xi \sim {1\over G m^3 N}&\sim&{ 8 \pi (10^{18}\,\mbox{GeV})^2\over (10^{-5}\,\mbox{eV})^3\times 10^{61}}\\
&\sim&\mbox{km}
\eea
On the other hand, the background mess of non-condensed scalar field can have a larger correlation length before galaxy formation.
However, inside galaxies, this scale $\sim\,\mbox{km}$ is within an order of magnitude or so of the de Broglie wavelength of virialized axions. 

As an upper limit, we note that according to
equation~(\ref{Jeans}), at the time of the QCD phase transition
the Jeans length was shorter than the Hubble length.  Roughly,
\bea
{k_J \over H_{\rm QCD}} &\sim& \left[ {2 m^3 n_a \over \mpl^2}
\right]^{1/4} {\mpl \over T_{\rm QCD}^2} 
\sim \left[ {2 m^2 T_{\rm eq} T_{\rm QCD}^3 \over \mpl^2}
\right]^{1/4} {\mpl \over T_{\rm QCD}^2} \nonumber \\
&\sim& {m^{1/2} T_{\rm eq}^{1/4} \mpl^{1/2} \over T_{\rm
QCD}^{5/4}}  \nonumber \\
&\sim& {(10^{-5} \, {\rm eV})^{1/2} (0.1\, {\rm eV})^{1/4}
(10^{18}\,{\rm GeV)}^{1/2} \over (100\,{\rm MeV})^{5/4}}
\nonumber \\
&\sim& 10 
\eea
Thus, immediately after the QCD phase transition, we expect
correlations up to the Hubble length, but the correlations with
wavelengths between the Jeans length and the Hubble length will
start to disappear, as perturbations grow. 
Equation~(\ref{Jeans}) shows that the Jeans length grows as
$a^{3/4}(t)$, so in comoving coordinates it shrinks with time. 
Thus, correlations with comoving wavelengths larger than the
Hubble length at the QCD phase transition can never form, since
causality forbids such correlations before the QCD phase
transition (assuming that the Peccei-Quinn phase transition
occurs after inflation), and afterward these wavelengths are
always larger than the Jeans length.  Thus, the comoving
correlation length cannot possibly exceed the Hubble length at
the QCD phase transition, which when scaled to today, is only on
the order of
\bea
\xi_{\rm rescaled\ Hubble} &\sim& H_{\rm QCD}^{-1} {T_{\rm QCD}
\over T_0} \sim {\mpl \over T_{\rm QCD} T_0} \nonumber \\
&\sim& {(10^{18} \, {\rm GeV}) \over (100\, {\rm MeV}) (10^{-4}\,
{\rm eV}) } \nonumber \\
&\sim& \hbox{light-year} 
\label{rescaled-Hubble}
\eea
which is much less than galactic scales. Thus, there does not
appear to be any mechanism for axion thermalization to lead to a
cosmologically large, or galactic scale, correlation length. A
full analysis of the production of these Bose stars requires a
full simulation, which is the topic of ongoing work.

We finish this section with a comment on the mass of these Bose stars.
Based on the above estimates, the typical mass is
\bea 
M=Nm&\sim& 10^{61}\times 10^{-5}\,\mbox{eV}\\
&\sim& 10^{-10}\,M_{\rm sun}
\eea 
This estimate is very close to the maximum possible mass of a stable QCD-axion star, about $10^{19}$ kg $\approx 5 \times 10^{-12}\,M_{\rm sun}$, that was found in Ref.~\cite{Eby:2014fya}. So they are much lighter than solar masses. Such light/low density objects are unlikely to have cosmological or galactic consequences. Furthermore, they may be outside the range of microlensing searches.

\bigskip
\section{Summary and Discussion}\label{Conclude}

In this paper, we investigated the idea that axion dark matter gravitationally thermalizes to form a Bose-Einstein condensate (BEC) with a long-range correlation length -- an intriguing idea that has a unique observational signature \cite{Sikivie:2009qn,Erken:2011dz}.  We treated the axion using a generic low-mass non-relativistic scalar field theory and studied its equilibrium behavior. While a BEC of long-range correlation can form from repulsive interactions, we showed that the homogeneous, or nearly homogeneous, condensate configuration is unstable against collapse when attractive interactions are included. Hence the state of long-range correlation is not the entropically favored equilibrium configuration. Instead the axions try to form a different type of BEC, namely clumps, either solitonic for self-interactions or Bose stars for self-gravity. The full state of the axions would be some complicated configuration of many BEC clumps that struggles to achieve true thermal equilibrium, and phases of distant clumps will tend to be uncorrelated. The correlation length should be rather small and not of cosmological significance. 

Our analysis applies to the QCD axion, which only has attractive interactions, and also to any other type of scalar dark matter candidate with attractive interactions in the high density/occupancy regime. At late times, the dominant interaction is ordinarily given by gravity, which of course is universally attractive. This means our analysis is very general. We worked in the classical field theory approximation, which is appropriate in this limit. Indeed we showed that classical fields can exhibit a phase transition to a BEC. We noted that Bose-Einstein condensation is very quantum mechanical from the particle perspective, but very classical from the field perspective.
If one moves to another regime of low occupancy, or low coherence, wherein the system is poorly approximated by the classical field theory, then other behavior would be possible. For example, if one passes to the classical particle phase space description, then the collision term in the Boltzmann equation does not depend on the sign of the coupling. The equilibrium behavior indicated by the standard Boltzmann equation therefore misses the essential equilibrium behavior of the field theory.

Ongoing work includes a full nonlinear simulation of the field theory to analyze the production of solitons and especially Bose stars (the latter is expected to be much more important at late times). Earlier work along these lines includes Ref.~\cite{Kolb:1993hw,Kolb:1993zz}.
This will also help to provide an understanding of the approach to equilibrium or otherwise. One might see a form of ``quasi-equilibrium" \cite{Khlebnikov:1999qy} wherein the clumps form and evaporate and so on. 

The qualitative difference in the size of the correlation length,  between attractive and repulsive interactions, should carry over to many more bosonic dark matter models. For example, in the string landscape it is possible to have many light axions \cite{Svrcek:2006yi,Arvanitaki:2009fg}. These should typically also have attractive self-interactions, so we expect similar conclusions to that of the QCD axion. One could also investigate scalar dark matter models, not motivated by axions, wherein the couplings are repulsive. 
%Or more general models with new long-range interactions that are repulsive. 
In these cases, the generation of long-range correlations is feasible, although highly parameter dependent. This could conceivably lead to novel new galactic behavior. Work on these subjects is ongoing.

\bigskip
\begin{center}{\bf Acknowledgments}\end{center}

%\bigskip

We would like to thank Niayesh Afshordi, Anthony Aguirre, Edmund Bertschinger, Jolyon Bloomfield, Latham Boyle, Alberto Diez, David Kaiser, Johanna Karouby, Emanuel Katz, Vinothan Manoharan, John Moffat, Sonia Paban, Evangelos Sfakianakis, Pierre Sikivie, and Frank Wilczek for helpful discussions. This work is supported by the U.S. Department of Energy under cooperative research agreement Contract Number DE-SC00012567. CP is supported by the Dr. Martin Luther King, Jr. Visiting Professors and Scholars program at M.I.T.

%\end{document}

\appendix
\section{Including FRW Expansion}\label{Bgd}

The relativistic action in a flat FRW background is
\beq
\mathcal{L}=a^3\left[{1\over2}\dot\phi^2-{1\over 2}{(\nabla\phi)^2\over a^2} -{1\over2}m^2\phi^2-{\lambda\over4!}\phi^4\right]
\eeq
where the scale factor is determined by the Friedmann equation
\beq
H^2={8\pi G\over3}\rho_{\rm tot}
\eeq
Passing to the non-relativistic field $\psi$ and ignoring rapidly varying terms gives
\beq
\mathcal{L}=a^3\left[{i\over2}(\dot\psi\psi^*-\psi\dot\psi^*)-{1\over2m}{\nabla\psi^*\!\cdot\!\nabla\psi\over a^2}-{\lambda\over 16m^2}(\psi^*\psi)^2\right]
\eeq
The corresponding classical equation of motion is
\beq
{i\over a^{3/2}}\partial_t(a^{3/2}\psi)=-{1\over2m}{\nabla^2\psi\over a^2}+{\lambda\over 8m^2}|\psi|^2\psi
\eeq

Including  gravity leads to the following pair of equations
\bea
{i\over a^{3/2}}\partial_t(a^{3/2}\psi) & = & -{1\over2m}{\nabla^2\psi\over a^2}+{\lambda\over 8m^2}|\psi|^2\psi+m\phi_N\psi\,\,\,\,\,\,\,\,\\
\nabla^2\phi_N & = & 4\pi Ga^2 \left(m|\psi|^2-\bar{\rho}\right)
\eea
where we have removed the background density in the source for the Newtonian potential $\phi_N$.

In order to solve the above equations we can make several simplifications.
Firstly, due to redshifting, the $\lambda\,\phi^4$ contact interaction is negligibly small at late times, so we will ignore it here. Secondly, we will linearize around a coherent homogeneous background as usual.
The solution for $\psi_c$ is
\beq
\psi_c(t)\propto {1\over a^{3/2}}
\eeq
The linearized equations of motion for the perturbations are
\bea
i\,\dot\dPsi &= & -{1\over 2ma^2}\nabla^2\dPsi+m\,\phi_N\\
\nabla^2\phi_N &=& 4\pi G m a^2 n_0(\dPsi+\dPsi^*)
\eea
Fourier transforming and then eliminating $\phi_N$ as before leads to 
\beq
i\,\dot\dPsi_k={k^2\over 2ma^2}\dPsi_k-{3\over 2}m\Omega_a{H^2a^2\over k^2}(\dPsi+\dPsi^*_k)
\eeq
where $\Omega_a=m\, n_0/\rho_{\rm tot}$.
Breaking up $\dPsi$ into real and imaginary as $A+iB$ and then eliminating $B$, we obtain
\beq
\ddot A_k+2 H\dot A_k-{3\over 2}\Omega_a H^2 A_k +\left(k^2\over 2 ma^2\right)^{\!2}\! A_k = 0
\eeq
The first three terms are the ``usual" terms one obtains for the growth of fluctuations in linearized theory of cold dark matter (CDM). The last term is a type of quantum pressure that arises from tracking the de Broglie wavelength of the axion. One can define a critical wavenumber where the pressure term balances the gravitation term (the Jeans wavenumber). It is given by
\beq
{k_J\over a} = (6\,\Omega_a)^{1/4}\sqrt{ H m}
\eeq
and coincides with the Jeans wavenumber of equation~(\ref{Jeans}) that we found in the absence of expansion.

For $k\ll k_J$ we can ignore the pressure term and we recover the usual equation for CDM. Its solutions are well known:
\bea
&&A_k\sim \log(a),\,\,\,B_k\propto a^0,\,\,\,\,\,\mbox{radiation era}\\
&&A_k\propto a, \,\,\, B_k\propto a^{3/2},\,\,\,\,\,\,\,\,\,\,\,\,\mbox{matter era}
\eea

For $k\gg k_J$ we are in the pressure dominated regime, dominated by oscillations. Putting in numbers, as in equation~(\ref{rescaled-Hubble}), we find that this regime corresponds to very small scales, probably irrelevant to the claims of Ref.~\cite{Erken:2011dz}.

\end{document}